\documentclass{iopart}
\usepackage[dvips]{graphicx}
\usepackage{amssymb}

\newcommand{\text}[1]{\hbox{\scriptsize\rm #1}}
\newcommand{\un}[1]{\mathrm{\,#1}}
\newcommand{\IJMP}{{\it Int. J. Mod. Phys.}}
\newcommand{\PRpt}{{\it Phys. Rep.}}
\begin{document}
\title{Progress on stochastic background search codes for LIGO}
\author{John T Whelan\dag,
  Warren G Anderson\dag, Martha Casquette\dag, Mario C D\'{\i}az\dag, Ik
  Siong Heng\ddag, Martin McHugh\S, Joseph D Romano\dag, Charlie W
    Torres Jr\dag, Rosa M Trejo\dag\footnote[4]{Present Address:
      Materials Science Program, Vanderbilt University, 303 Olin Hall,
      Nashville, Tennessee 37240, USA} and Alberto Vecchio\P$^+$}
\address{{\dag} Department of Physics and Astronomy, The University of Texas
  at Brownsville, Brownsville, Texas 78520, USA}
\address{{\ddag} Department of Physics and Astronomy, Louisiana State
  University, Baton Rouge, Louisiana 70803, USA}
\address{{\S} Department of Physics, Loyola University, New Orleans,
  Louisiana 70118, USA}
\address{{\P} School of Physics and Astronomy, The University of Birmingham,
  Edgbaston, Birmingham B15 2TT, United Kingdom}
\address{$^+$ Max-Planck-Institut f\"ur Gravitationsphysik, 
  Albert-Einstein-Institut, Am~M\"uhlenberg~1, 14476~Golm,~Germany}

\begin{abstract}
  One of the types of signals for which the LIGO interferometric
  gravitational wave detectors will search is a stochastic background
  of gravitational radiation.  We review the technique of searching
  for a background using the optimally-filtered cross-correlation
  statistic, and describe the state of plans to perform such
  cross-correlations between the two LIGO interferometers as well as
  between LIGO and other gravitational-wave detectors, in particular
  the preparation of software to perform such data analysis.
\end{abstract}

\submitto{\CQG}
\ead{whelan@oates.utb.edu}
\maketitle

\section{Data Analysis Techniques}\label{sec:techniques}

In this section we review the data analysis technique to be used to
detect a stochastic background of gravitational radiation.  More
details can be found in \cite{Allen:1997,Allen:1999}.

\subsection{Definitions}

First, we limit attention to backgrounds which are cosmological in
origin and thus can be assumed to be isotropic, unpolarized, Gaussian,
and stationary.  Subject to these assumptions, the stochastic
gravitational-wave (GW) background is completely described by its power
spectrum.  It is conventional to express this spectrum in terms of the
GW contribution to the cosmological parameter
$\Omega=\rho/\rho_{\text{crit}}$:
\begin{equation}
  \label{eq:omegagw}
  \Omega_{\text{GW}}(f)=\frac{1}{\rho_{\text{crit}}}
  \frac{d\rho_{\text{GW}}}{d\ln f}=\frac{f}{\rho_{\text{crit}}}
  \frac{d\rho_{\text{GW}}}{df}
  \ .
\end{equation}
Note that $\Omega_{\text{GW}}(f)$ has been constructed to be
dimensionless, and represents the contribution to the overall
$\Omega_{\text{GW}}$ per \emph{logarithmic} frequency interval.  In
particular, it is \emph{not} equivalent to $d\Omega_{\text{GW}}/df$.
Note also that since the critical density $\rho_{\text{crit}}$, which
is used in the normalization of $\Omega_{\text{GW}}(f)$, is
proportional to the square of the Hubble constant $H_0$ \cite{Kolb:1990},
it is convenient to work with $h_{100}^2\Omega_{\text{GW}}(f)$, which
is independent of the observationally determined value of
$h_{100}=\frac{H_0}{100 \un{km}/\un{s}/\un{Mpc}}$.

\subsection{Cross-Correlation}

Since a stochastic signal is by definition random, it is impractical
to look for one in the output of a single gravitational wave detector
\cite{Christensen:1992}.
However, the effects of such a signal can be detected by
cross-correlating the outputs of two independent detectors.  If the
output $h_1$ ($h_2$) of the first (second) detector consists of a term
$s_1$ ($s_2$) due to stochastic gravitational waves and a term $n_1$
($n_2$) due to instrument noise, and the noise in each instrument is
assumed to be uncorrelated both with the GW signal and
the noise in the other instrument, the only surviving term in a time
averaged correlation $\langle h_1 h_2 \rangle$ is the term $\langle
s_1 s_2 \rangle$ due to the gravitational wave background.

In practice, one defines a \emph{cross-correlation statistic}
\begin{equation}
  Y_Q
  =
  \int dt_1\, dt_2\, {h_1(t_1)}\, {Q(t_1-t_2)}
  {h_2(t_2)}
  \label{eq:CCstat}
  =
  \int df\,{\widetilde{h}_1^*(f)}\, {\widetilde{Q}(f)}\,
  {\widetilde{h}_2(f)}
\end{equation}                                
which is weighted by a filter $\widetilde{Q}(f)$.  In the presence of
a stochastic gravitational wave signal, both the mean $\mu$ and
variance $\sigma^2$ of the cross-correlation statistic will grow
linearly with time, so the signal-to-noise ratio $\mu/\sigma$ will
grow as the square root of the observation time.

Given a GW background spectrum and a pair of detectors, the
signal-to-noise ratio is maximized by using the \emph{optimal filter}
\begin{equation}
  \label{eq:optfilt}
  \widetilde{Q}(f)\propto 
  \frac{f^{-3}\Omega_{\text{GW}}(f)\gamma_{12}(f)}
  {P_1(f)P_2(f)}
\ .
\end{equation}
The denominator, containing the power spectral densities $P_{1,2}(f)$
of the noise in the two detectors, serves to suppress the
contributions to the cross-correlation statistic from frequencies
where one or both detectors are ``noisy'' and most correlations are
therefore likely to be accidental.  The numerator represents the
average unweighted cross-correlation between the outputs of two
detectors, and depends on both the spectrum $\Omega_{\text{GW}}(f)$ of
the expected gravitational wave background and the locations and
orientations of the two detectors.  The latter is described by the
\emph{overlap reduction function} $\gamma_{12}(f)$
\cite{Flanagan:1993}.

\begin{figure}[htbp]
  \centering
  \includegraphics[height=3.5in]{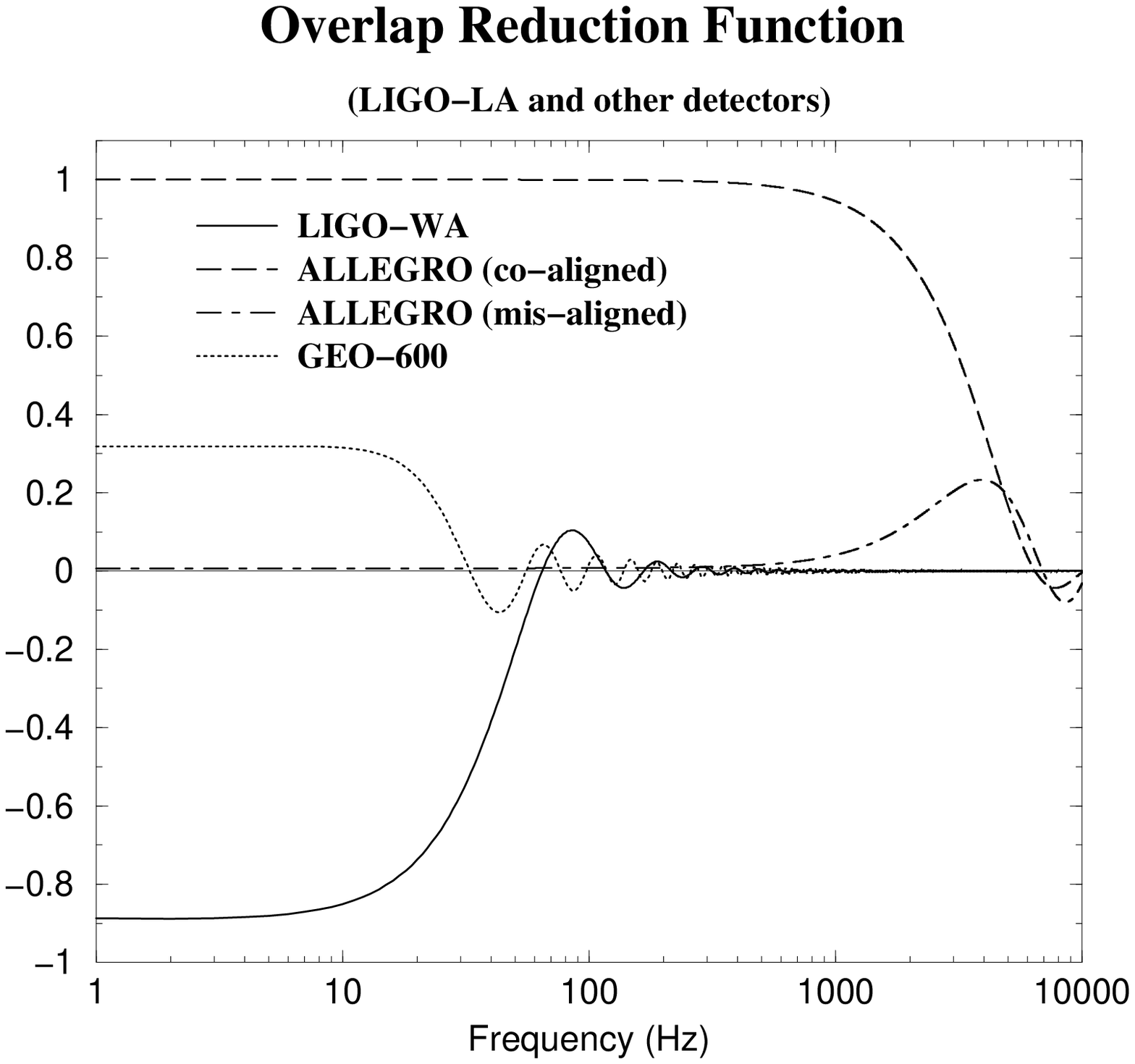}
  \caption{The overlap reduction function for several pairs of
    gravitational wave detectors.  In each case, one of the detectors
    is the LIGO-LA site in Livingston, Louisiana.  The curve labelled
    ``LIGO-WA'' shows the overlap with the site in Hanford,
    Washington; the one labelled ``GEO-600'' is for LIGO-LA and the
    GEO-600 site in Hannover, Germany, and the curves labelled
    ``ALLEGRO'' refer to the ALLEGRO resonant bar detector in Baton
    Rouge, Louisiana.  The ALLEGRO experimental setup allows for the
    orientation of the detector to be changed.  ``ALLEGRO
    (co-aligned)'' shows the overlap reduction function when ALLEGRO
    is oriented approximately parallel to one of the arms of LIGO-LA;
    ``ALLEGRO (mis-aligned)'' corresponds to an orientation $45^\circ$
    away from this.}
  \label{fig:overlap}
\end{figure}

The overlap reduction function is equal to unity for the case of a
pair of interferometers (IFOs) at the same location with their arms
aligned, and is suppressed as the detectors are rotated out of
alignment or separated from one another.  The frequency dependence
comes about for the following reason: if the wavelength of a wave is
comparable to or smaller than the separation between two detectors,
the detectors will see different phases of the wave at the same time,
and this phase difference will depend on the direction of propagation
of the wave.  Since the stochastic GW background is assumed to be
isotropic, averaging over different propagation directions suppresses
the sensitivity of a pair of detectors to high-frequency waves.  For 
example, a wave whose wavelength is twice the distance between the
two detectors will drive them $180^\circ$ out of phase if it travels
along the line separating them, but \emph{in} phase if its direction
of propagation is perpendicular to this line.
Figure~\ref{fig:overlap} shows the overlap reduction functions for
several detectors of interest.

\subsection{Cross-Correlation Spectrum}

In early data analysis applications, cross-correlated noise is likely
to produce considerable spurious contributions to the integral
(\ref{eq:CCstat}).  A useful diagnostic tool will thus be the
\emph{cross-correlation spectrum}
\begin{equation}
  \label{eq:CCspec}
  Y(f)
  = {\widetilde{h}_1^*(f)}\, {\widetilde{Q}(f)}\,
  {\widetilde{h}_2(f)}
\ ,
\end{equation}
which is simply the integrand of (\ref{eq:CCstat}); this may enable us
to see directly the impact of these cross-correlated noise sources on
the cross-correlation statistic.

\section{Setting upper limits with LIGO}

A two-week engineering run for the LIGO IFOs is
planned for around the end of 2001, involving the kilometer-scale IFOs
in Livingston, Louisiana \cite{LLO} and Hanford, Washington
\cite{LHO}.  The 600 meter GEO IFO \cite{GEO} in Hannover, Germany and
the ALLEGRO \cite{ALLEGRO} resonant bar detector at Louisiana State
University are also planning to operate during the same two-week
period.  Four groups have been formed to use the data taken from LIGO
and the other instruments to set upper limits on various types of GW
signals, including stochastic signals \cite{stochUL}.  The stochastic
group plans to set an upper limit on the strength of a stochastic
background, assuming it has the form
$\Omega_{\text{GW}}(f)=\textrm{constant}$, with the goal of improving
on the existing best upper limit of $\Omega_{\text{GW}}(f)\lesssim 60$
\cite{Astone:2000}.

The roles to be played by various pairs of detectors in this effort
are largely driven by their separation and alignment relative to one
another, as quantified by the overlap reduction function (see
Figure~\ref{fig:overlap}).

\subsection{Correlations between LIGO-LA and LIGO-WA}

The distance between the two LIGO sites is approximately 3000 km,
which makes the light-travel time between the two sites about 10 ms.
Thus the two sites are separated by half a wavelength for waves with a
frequency of 50~Hz, and as shown in Figure~\ref{fig:overlap} the
overlap reduction function first crosses zero at a slightly higher
frequency.  So the overlap reduction function limits the sensitivity
of this pair of detectors at high frequencies; below around 40~Hz, the
seismic noise in the detectors will squash the optimal filter
(\ref{eq:optfilt}).  The net effect (as illustrated in figure~21 of
\cite{Allen:1999}) is that for LIGO initial design sensitivity, most
of the support of the optimal filter lies between 50 and 250~Hz.

\subsection{Correlations between LIGO-LA and ALLEGRO}

The ALLEGRO bar detector is far closer to the LIGO Livingston site
than the LIGO Hanford site is, with only about 40 km separating the
two and a ``half-wavelength'' frequency of 3750~Hz.  Thus the
observing geometry allows for observations of correlations out to much
higher frequencies.  On the other hand, the sensitivity of ALLEGRO is
concentrated in two narrow frequency bands in the vicinity of 900~Hz,
so correlations between ALLEGRO and LIGO Livingston probe a different
part of the frequency domain than correlations between the two LIGO
detectors.\footnote{Another consequence of the proximity between the
  two detectors is that there is likely to be a lot of
  cross-correlated noise; a method \cite{Finn:2001} has been proposed
  to account for this noise by measuring the cross-correlation for
  different alignments of the ALLEGRO bar.}

\subsection{Correlations between LIGO-LA and GEO-600}

Since the GEO-600 detector is rather distant from the LIGO sites (over
7500~km from Livingston, corresponding to a ``half-wavelength''
frequency of only 20~Hz), the small overlap reduction function will
render the GEO-600/LIGO-LA pair (the better of the two) considerably
less sensitive to gravitational waves than the LIGO-LA/LIGO-WA pair.
The primary interest in performing this correlation is thus the
information it will provide about cross-correlated noise rather than a
contribution to the upper limit on stochastic background strength.

\section{Data Analysis Routines}

\subsection{Routines in the LIGO Numerical Algorithms Library (LAL)}

The data analysis technique described in Section~\ref{sec:techniques}
will be implemented within the LIGO data analysis system (LDAS)
\cite{ldas} using C routines from the LIGO numerical Algorithms
Library (LAL) \cite{lal}.  We have written and tested LAL routines to
perform the various parts of the analysis (calculating the overlap
reduction function, constructing the optimal filter, etc.).

Care has been taken to make these routines general enough to be
applied to both IFO data (e.g., from LIGO and GEO) and data
from resonant bar detectors such as ALLEGRO.  Two major issues have
required some care in this regard:

First, the treatment of detector geometry (used in constructing the
optimal filter) needed to be general enough to describe both
interferometric and resonant detectors.  This was accomplished by
defining a data structure within LAL which described an idealized
earthbound detector in terms of its location and tensor response to
gravitational waves \cite{LALDetector}.

Second, since the sensitivity band of ALLEGRO is at a relatively high
frequency compared to its bandwidth, its gravitational wave signal is
heterodyned before being discretely sampled \cite{hetero}.  By
multiplying the time-domain signal by a complex exponential
oscillating at a base frequency, one effectively shifts the frequency
band represented in the discrete signal (whose full width is equal to
the sampling frequency) so that it is centered at the base frequency
rather than at DC (0~Hz).  To allow for this, the LAL routines had to
be written to deal with complex as well as real time series.

\subsection{Driver Routines in LALWrapper}

The interface between the C$++$ LDAS environment and the LAL C library
is known as LALWrapper \cite{LALWrapper}.  LALWrapper contains a
number of dynamically-linked shared objects which can be used to
``drive'' various search algorithms in LAL.  We have written two
LALWrapper shared objects: \texttt{libldasstochastic.so} calculates
the cross-correlation spectrum (\ref{eq:CCspec}) between two
interferometric detectors, and \texttt{libldasstochasticbar.so} does
the same for correlations between an IFO and a resonant bar detector.
(Eventually, we plan to integrate the functionality into a single,
generalized search engine.)

The general behavior of LALWrapper code is to execute parallel
searches on one or more nodes of a Beowulf cluster \cite{beowulf},
with each ``slave'' node reporting to the ``master'' node at least ten
times.  Both stochastic background search engines use the following
algorithm to calculate cross-correlation spectra for a sequence of
short consecutive time intervals:
\begin{enumerate}
\item Equal-length stretches of data from a pair of detectors are
  input, along with power spectra and response functions, and some
  search parameters.
\item The data streams are each divided into ten or more
  shorter-length segments.
\item An optimal filter is constructed using the auxiliary inputs and
  parameters describing the choice of detectors, etc.
\item In turn, each corresponding pair of data segments is
  Fourier-transformed and the cross-correlation spectrum calculated
  using the optimal filter.
\end{enumerate}

Enhancements to be made for the scientific data runs which will begin
in 2002 include: (i) Rather than constructing a single optimal filter
based on the $\Omega_{\text{GW}}(f)=\textrm{constant}$ model, we will
choose a set of points in the parameter space of stochastic background
models\cite{filtergrid} and filter the data in parallel with a
``grid'' of optimal filters, one optimized according to
(\ref{eq:optfilt}) for each $\Omega_{\text{GW}}(f)$ model
\cite{Maggiore:2000}; (ii) To set an upper limit or make a measurement
of the strength of the stochastic background, we will calculate the
cross-correlation statistic for the whole data stream(s) by
integrating the cross-correlation spectrum for a given segment over
frequency and then adding the contributions from all the segments.

\subsection{Mock Data Challenge}

The shared objects \texttt{libldasstochastic.so} and
\texttt{libldasstochasticbar.so} were among the data analysis routines
tested at the Burst-Stochastic Mock Data Challenge\cite{MDC}, held at
the Massachusetts Institute of Technology 4-10 September 2001.  
The shared object for IFO-IFO correlations was tested with trivial and
non-trivial synthetic data and produced the expected results in each
case; it was also used to analyze 15 minutes of data taken by the two
LIGO IFOs during a recent engineering run, to verify that it could do
so without failing.  The IFO-bar shared object, being slightly less
mature, was not tested as extensively.  However, some of the tests
which were run produced unexpected results which appear to be due to
the use of single precision arithmetic in several LAL routines.  The
finer frequency resolution required by the sharp spectral features in
the bar response function may require the use of double precision.

\ack


We would like to thank our colleagues in LIGO and the larger
gravitational wave community who have made this work possible,
especially: B.~Allen and the authors of the GRASP routines from which
many LAL routines are derived; S.~Drasco and \'{E}.~Flanagan, who
wrote early versions of several LAL stochastic background routines;
LSC Software Co\"{o}rdinator A.~Wiseman and LAL Librarian
J.~Creighton, without whom LAL would not exist; the other members of
the site structure development team including P.~Brady, D.~Chin,
J.~Creighton, C.~Cutler, K.~Riles and A.~Lazzarini, who came up with
the idea of using the response tensor to describe a generic GW
detector; our fellow participants in the Burst-Stochastic MDC
including A.~Searle on the stochastic side and S.~Finn, E.~Daw,
S.~Marka, P.~Saulson and J.~Zweizig on the burst side and especially
E.~Katsavouinidis and J.~Sylvestre for hosting and K.~Blackburn,
P.~Charlton, P.~Shawhan, M.~Barnes, P.~Ehrens, M.~Lei and I.~Salzman
for providing LDAS support; and finally the other members of the
Stochastic Sources Upper Limits Group, including S.~Bose,
N.~Chistensen, R.~Drever, P.~Fritschel, J.~Giaime, W.~Hamilton,
W.~Johnson, M.~Landry, T.~Nash, A.~Ottewill, B.~Whiting and R.~Weiss.
This work was supported by the National Science Foundation under
grants PHY-9981795 (UTB) and PHY-9970742 (LSU), and by NASA contract
JPL1219731 (UTB).
Figure~\ref{fig:overlap} was made using slightly modified versions of
LAL routines \cite{lal}.

\end{document}